\documentclass{appolb}
\usepackage{graphicx}
\usepackage{multirow}
\usepackage{url}
\begin{document}
\title{B(E2$\uparrow$) strength in $^{36,38}$Ca and in mirror nuclei $^{36}$S, $^{38}$Ar.%
\thanks{Presented at the 57th Zakopane Conference on Nuclear Physics, {\it Extremes of the Nuclear Landscape}, Zakopane, Poland, 25 August–1 September, 2024.}%
}
\author{Jacek Oko{\l}owicz
\address{Institute of Nuclear Physics, Polish Academy of Sciences, Radzikowskiego 152, PL-31342 Krak{\'o}w, Poland}
\\[3mm]
{Marek P{\l}oszajczak
\address{Grand Acc\'el\'erateur National d'Ions Lourds (GANIL), CEA/DSM - CNRS/IN2P3, BP 55027, F-14076 Caen Cedex 05, France}
}
}
\maketitle
\begin{abstract}
Recently, $B(E2,0^+_1 \rightarrow 2^+_1)$ transition strength have been measured in $^{36}$Ca and $^{38}$Ca. Surprisingly, the measured value in $^{36}$Ca: $B(E2\uparrow)$ = 131(20) $e^2$fm$^4$, is significantly larger than in $^{38}$Ca, where $B(E2\uparrow)$ = 101(11) $e^2$fm$^4$, whereas an opposite tendency of $B(E2)$ values is seen in the mirror nuclei $^{36}$S and $^{38}$Ar. The resonance $2^+_1$ in $^{36}$Ca lies 465 keV above the proton emission threshold and its description requires inclusion of the coupling to the continuum. In this work, we are analyzing  $B(E2\uparrow)$ values in $^{36}$Ca, $^{36}$S, $^{38}$Ca, and $^{38}$Ar using the real-energy continuum shell model, the so-called shell model embedded in the continuum, in the $(1s_{1/2}\, 0d_{3/2}\, 0f_{7/2}\, 1p_{3/2})$ model space with the monopole adjusted effective interaction ZBM-IO.
\end{abstract}
  
\section{Introduction}
$B(E2,0^+_1 \rightarrow 2^+_1)$ transition strength provides a sensitive test of the ground state and the first $2^+$ state wave functions in even-even nuclei. This experimental observable is also a useful indicator of the shell evolution and the variation of the nucleon-nucleon correlations. It also indicates how well the mirror symmetry is satisfied.
The recent measurement of $B(E2)$ values in $^{36,38}$Ca~\cite{PhysRevC.107.034306,PhysRevC.108.L061301} provided astonishingly different results as compared to the results in mirror nuclei $^{36}$S, $^{38}$Ar. 

$^{36}$Ca has the proton shell $Z=20$ and the neutron subshell $N=16$  closed~\cite{PhysRevLett.100.152502}, whereas $^{38}$Ca lies in-between N=16 and N=20 shell closures. Indeed, the first excited state in $^{36}$Ca is the $0^+_2$ state, as one might expect in closed shell nuclei. The $2^+_1$ state in this nucleus is one- and two-proton unbound~\cite{APP2008}. It lies 465 keV above the one-proton emission threshold~\cite{399d5bb09b5d11dcbee902004c4f4f50} and $\sim$230 keV above the $0^+_2$ state~\cite{PhysRevLett.129.122501}. It couples to the ground states of $^{35}$K and $^{34}$Ar in $\ell = 0$ and $\ell = 2$ partial waves, respectively. Since there is no centrifugal barrier for the one-proton (1p) decay, and the available phase space for 1p decay is significantly larger than for two-proton (2p) decay therefore, one expects that the 1p emission dominates the decay of this resonance.

 The mirror nucleus $^{36}$S is well bound. Contrary to $^{36}$Ca, the first excited state is $2^+_1$, 6.6 MeV below the one-neutron emission threshold and  $\sim$50 keV below the $0^+_2$ state. This implies that the influence of continuum coupling on the structure of $2^+_1$ states in $^{36}$Ca and $^{36}$S can be different.  Indeed, the difference of $2^+_1$ excitation energies in these nuclei is  $\sim$250 keV, whereas  the corresponding energy difference in well bound nuclei $^{38}$Ca and $^{38}$Ar is only $\sim$50 keV.
 
 Coupling to the continuum may also play a role in the anomalous \\ $B(E2,0^+_1 \rightarrow 2^+_1)$ transitions probabilities in mirror pairs of nuclei: $^{36}$Ca and $^{36}$S, for which the $B(E2\uparrow)$ transition probabilities are $131\pm 20$~$e^2$~fm$^4$~\cite{PhysRevC.107.034306,PhysRevC.108.L061301} 
 and $89\pm 9$~$e^2$~fm$^4$~\cite{399d5bb09b5d11dcbee902004c4f4f50}, respectively, and  $^{38}$Ca and $^{38}$Ar for which experimental values are $101\pm 10$~$e^2$~fm$^4$~\cite{PhysRevC.107.034306,PhysRevC.108.L061301}  and $125\pm 4$~$e^2$~fm$^4$~\cite{PRITYCHENKO201773}, respectively. Whereas $B(E2\uparrow)$ in $^{36}$Ca is larger than in $^{38}$Ca, the tendency is opposite in mirror nuclei  and $B(E2\uparrow)$ is larger in $^{38}$Ar   than in $^{36}$S. 
 
 In this work, we shall apply the shell model embedded in the continuum (SMEC)~\cite{Okolowicz2003,bennaceur_2000,rotureau_2006}, the real-energy continuum shell model, to analyze the $B(E2,0^+_1 \rightarrow 2^+_1)$ transition strength. The spectrum of $^{36}$Ca in SMEC was discussed previously in Ref.~\cite{APP2008}. Brief presentation of SMEC and the Hamiltonian is given in Sec.~\ref{sec:theory}. Discussion of results and a short summary is in Sec.~\ref{sec:discussion}.

\section{Shell model embedded in the continuum}
\label{sec:theory}
The SMEC has been extensively applied to calculate spectra and reactions for bound, weakly bound and unbound nuclear states~\cite{Okolowicz2003,bennaceur_2000,rotureau_2006}. Here we will present only essential  features of this model and more details can be found in Refs.~\cite{Okolowicz2003,rotureau_2006}. 

Hilbert space in the SMEC is divided into orthogonal subspaces ${\cal Q}_{0}$, ${\cal Q}_{1}$, ${\cal Q}_{2}$, $\dots$ containing 0, 1, 2, $\dots$ particles in the scattering continuum. Since the two-proton decay provides only a tiny contribution to the total decay width of the $2^+_1$ resonance in $^{36}$Ca, we shall restrict our discussion to the simplest version of SMEC with the two subspaces  ${\cal Q}_{0}$, ${\cal Q}_{1}$ only. An open quantum system  description of ${\cal Q}_0$  includes couplings to the environment of decay channels through the energy-dependent effective Hamiltonian:
\begin{equation}
{\cal H}(E)=H_{{\cal Q}_0{\cal Q}_0}+W_{{\cal Q}_0{\cal Q}_0}(E),
\label{eq21}
\end{equation}
where $H_{{\cal Q}_0{\cal Q}_0}$ denotes the standard shell model (SM) Hamiltonian which describes internal dynamics, and 
\begin{equation}
W_{{\cal Q}_0{\cal Q}_0}(E)=H_{{\cal Q}_0{\cal Q}_1}G_{{\cal Q}_1}^{(+)}(E)H_{{\cal Q}_1{\cal Q}_0},
\label{eqop4}
\end{equation}
is the energy-dependent continuum coupling term, where $E$ is the scattering energy, $G_{{\cal Q}_1}^{(+)}(E)$ is the one-nucleon Green's function, and ${H}_{{Q}_0,{Q}_1}$ and ${H}_{{Q}_1{Q}_0}$ couple the subspaces ${\cal Q}_{0}$ with ${\cal Q}_{1}$. The channel state is defined by the coupling of one nucleon in the scattering continuum to the SM wave function of the nucleus  $(A-1)$. The SMEC eigenstates $|\Psi_{\alpha}^{J^{\pi}}\rangle$ of ${\cal H}(E)$ are the linear combinations of SM eigenstates $|\Phi_{i}^{J^{\pi}}\rangle$ of $H_{{\cal Q}_0{\cal Q}_0}$. 

As the $H_{{\cal Q}_0{\cal Q}_0}$ we take the ZBM-IO effective interaction which is defined in $(1s_{1/2}\, 0d_{3/2}\, 0f_{7/2}\, 1p_{3/2})$ model space~\cite{CAURIER2001240} with the two-body matrix elements of IOKIN interaction~\cite{PhysRevC.63.044316} and the modified $T=1$ cross-shell monopole terms ${\cal M}(1s_{1/2} 0f_{7/2})$, ${\cal M}(1s_{1/2} 1p_{3/2})$ which are adjusted to reproduce the low-lying states in the studied nuclei. It was argued in Ref.~\cite{PhysRevC.107.034306}, that the ZBM2 Hamiltonian~\cite{CAURIER2001240}, defined in the same model space ($1s_{1/2}\, 0d_{3/2}\, 0f_{7/2}\, 1p_{3/2}$),  provides best SM description of the $B(E2,0^+_1 \rightarrow 2^+_1)$ transitions probabilities in $^{36,38}$Ca. In the present calculation, we restrict the number of nucleon excitations from $(1s_{1/2} 0d_{3/2})$ to $(0f_{7/2} 1p_{3/2})$ to 2 nucleons.  

For the continuum-coupling interaction we take the Wigner-Bartlett contact force: 
$V_{12}=V_0 \left[ \alpha + \beta P^{\sigma}_{12} \right] \delta(\vec{r}_1-\vec{r}_2)$, 
where $\alpha + \beta = 1$, $P^{\sigma}_{12}$ is the spin exchange operator, and  the spin-exchange parameter takes a standard value $\alpha = 0.73$ \cite{Okolowicz2003}. 
The radial single particle wave functions (in ${\cal Q}_0$) and the scattering wave functions (in ${\cal Q}_1$) are generated by the average potential which includes the central Woods-Saxon term, the spin-orbit term, and the Coulomb potential. The radius and diffuseness of the Woods-Saxon and spin-orbit potentials are $R_0=1.27 A^{1/3}$~fm and $a=0.67$~fm, respectively. The strength of the spin-orbit potential is $V_{\rm SO}=6.7$ MeV for protons and $7.62$~MeV for neutrons. The Coulomb part is calculated for an uniformly charged sphere with the radius $R_0$.

\section{Discussion of results}
\label{sec:discussion}

\begin{table}[htb]
    \caption{The family of SMEC solutions for $^{36}$Ca and $^{38}$Ca. In the first two columns, the $T=1$ cross-shell monopoles (in MeV) are presented. Third column contains the continuum coupling strength $V_0$ (in MeV fm$^3$) which for each set of $T=1$ cross-shell monopoles in the first two columns is adjusted to reproduce the proton separation energy and energy of the $2^+_1$ state. Next columns contain the following sequence: $B(E2\uparrow)$ values (in units of $e^2$fm$^4$), and the ground state fractions $F_p(0)$, $F_p(2)$, $F_n(0)$ and $F_n(2)$ of proton/neutron parts excited from $(1s_{1/2} 0d_{3/2})$ to $(0f_{7/2} 1p_{3/2})$ shells. $B(E2\uparrow)$ values in this table have been calculated with the  effective charges: $e_p=1.236$, $e_n=0.409$.}
   \vskip 0.3truecm     
  \begin{center}
    \begin{tabular}{cccccccc}
    \hline
    \hline
   \multicolumn{2}{c}{${\cal M}$} &  &  \multicolumn{5}{c}{$^{36}$Ca}   \\
   $sf_{7/2}$ &  $sp_{3/2}$ & $V_0$ & $B(E2\uparrow)$ & $F_n(0)$ & $F_n(2)$ & $F_p(0)$ & $F_p(2)$  \\
    \hline
  -2.078 & -2.657  &  0    & 110  & 0.977 & 0.003 & 0.659 & 0.322  \\
  -2.137  & -2.554 & -61.3 &  116.7 & 0.978 & 0.003 & 0.635 & 0.346 \\
  -2.177  & -2.477   & -85   & 121 & 0.979 & 0.003 & 0.616 & 0.366   \\
  -2.227   & -2.377   & -113  & 126 & 0.980 & 0.003 & 0.588 & 0.395  \\
  -2.277  & -2.247  & -139  & 129 & 0.981 & 0.003 & 0.557 & 0.427  \\
   \hline
   \hline
 \multicolumn{2}{c}{${\cal M}$} &  &  \multicolumn{5}{c}{$^{38}$Ca}   \\
   $sf_{7/2}$ &  $sp_{3/2}$ & $V_0$ & $B(E2\uparrow)$ & $F_n(0)$ & $F_n(2)$ & $F_p(0)$ & $F_p(2)$  \\
    \hline
     -2.078 & -2.657  &  0 & 66.9 & 0.870 & 0.045 & 0.613 & 0.303 \\
   -2.137  & -2.554 & -61.3 & 82.5 & 0.873 & 0.045 & 0.580 & 0.337 \\
   -2.177  & -2.477   & -85   &  93.4 & 0.874 & 0.044 & 0.554 & 0.364 \\
  -2.227   & -2.377   & -113  &  108 & 0.877 & 0.042 & 0.519 & 0.400 \\
  --2.277  & -2.247  & -139  &  121 & 0.881 & 0.040 & 0.482 & 0.440 \\
   \hline  
   \hline
   \end{tabular}
   \end{center}
  \label{tab:smecres}
\end{table}
Table~\ref{tab:smecres} presents the continuum coupling strength $V_0$ and the corresponding $T=1$ cross-shell monopoles of the ZBM-IO interaction, which allow to describe proton separation energy, excitation energy of the $2^+_1$ state, and yield the $B(E2\uparrow)$ transition probabilities compatible with the data  within the experimental uncertainties for $^{36}$Ca and $^{38}$Ca simultaneously. One can see that the monopoles: ${\cal M}^{T=1}(1s_{1/2}\,1p_{3/2})$ and ${\cal M}^{T=1}(1s_{1/2}\,0f_{7/2})$, are anti-correlated and that the demanded criteria are satisfied by the whole family of solutions. 

Table~\ref{tab:smecres} shows also the fractions $F_n(0)$, $F_n(2)$, $F_p(0)$ and $F_p(2)$ of the proton and neutron parts in the ground state wave function which are excited from $(1s_{1/2} 0d_{3/2})$ to $(0f_{7/2} 1p_{3/2})$. The sum: $F_{n/p}(0)$+$F_{n/p}(2)$+$F_{np}(11)$, where $F_{np}(11)$ is the fraction corresponding to the simultaneous excitation of one proton and one neutron from $(1s_{1/2} 0d_{3/2})$ to $(0f_{7/2} 1p_{3/2})$, is normalized to 1. One may notice that the fraction $F_p(0)$ depends strongly on the monopoles ${\cal M}^{T=1}(1s_{1/2}\,1p_{3/2})$, ${\cal M}^{T=1}(1s_{1/2}\,0f_{7/2})$, and the continuum-coupling strength $V_0$, whereas $F_{np}(11)$ remains practically constant. Changes of $F_n(0)$ in the whole range of parameters are small. Moreover, $F_n(0)$ and $F_p(0)$ in $^{38}$Ca are significantly smaller than in $^{36}$Ca.

\begin{table}[htb]
    \caption{The family of SMEC solutions for mirror nuclei: $^{36}$S and $^{38}$Ar. For details see the caption of Table~\ref{tab:smecres}. }
   \vskip 0.3truecm     
   \begin{center}
    \begin{tabular}{cccccccc}
    \hline
    \hline
   \multicolumn{2}{c}{${\cal M}$} &  &  \multicolumn{5}{c}{$^{36}$S}   \\
   $sf_{7/2}$ &  $sp_{3/2}$ & $V_0$ & $B(E2\uparrow)$ & $F_n(0)$ & $F_n(2)$ & $F_p(0)$ & $F_p(2)$  \\
    \hline
  -2.112 & -1  &  0    &  89.0  & 0.697 & 0.288 & 0.982 & 0.003 \\
  -2.167  & -0.077   & -83 &  75.8 & 0.670 & 0.318 & 0.984 & 0.003 \\
  -2.183  & 0.423   & -96  & 71.7 & 0.662 & 0.326 & 0.985 & 0.003 \\
  -2.194   & 0.923   & -105.3  & 68.5 & 0.655 & 0.333 & 0.985 & 0.003  \\
  -2.207  &  1.923  & -116  & 64.7 & 0.648 & 0.341 & 0.986 & 0.003  \\
   \hline
    \hline
 \multicolumn{2}{c}{${\cal M}$} &  &  \multicolumn{5}{c}{$^{38}$Ar}   \\
   $sf_{7/2}$ &  $sp_{3/2}$ & $V_0$ & $B(E2\uparrow)$ & $F_n(0)$ & $F_n(2)$ & $F_p(0)$ & $F_p(2)$  \\
    \hline
   -2.112 & -1  &  0 & 126.4 & 0.601 & 0.324 & 0.883 & 0.043 \\
   -2.167  & -0.077  & -83 & 126.4 & 0.559 & 0.368 & 0.886 & 0.042 \\
   -2.183  & 0.423   & -96   &  125.0 & 0.546 & 0.382 & 0.887 & 0.041 \\
   -2.194  & 0.923  & -105.3  &  125.9 & 0.536 & 0.393 & 0.888 & 0.041 \\
   -2.207  &  1.923  & -116  &  125.6 & 0.522 & 0.407 & 0.889 & 0.040 \\
   \hline  
   \hline
   \end{tabular}
   \end{center}
  \label{tab:smecres-mir}
\end{table}
 Table~\ref{tab:smecres-mir} contains information on the family of monopole terms ${\cal M}^{T=1}(1s_{1/2}\,1p_{3/2})$, ${\cal M}^{T=1}(1s_{1/2}\,0f_{7/2})$, and the continuum-coupling strength $V_0$,  which allow to reproduce the neutron separation energy, excitation energy of the $2^+_1$ state, and provide the $B(E2\uparrow)$ values which are compatible with the experimental data in $^{36}$S and $^{38}$Ar.
 Changes of the ${\cal M}^{T=1}(1s_{1/2},0f_{7/2})$ monopole are small, while those of ${\cal M}^{T=1}(1s_{1/2},1p_{3/2})$ are large and  different from the corresponding monopole in $^{36}$Ca and $^{38}$Ca.
 Range of $F_p(0)$ variation in $^{36}$Ca ($^{38}$Ca) is significantly larger than the corresponding  variations of $F_n(0)$ in the mirror nuclei $^{36}$S ($^{38}$Ar). It is important to notice that $B(E2\uparrow)$ for $^{38}$Ar is almost constant in a large interval of parameters ${\cal M}^{T=1}(1s_{1/2}\,1p_{3/2})$ and $V_0$. This property has been used to tune the effective charges for all considered nuclei: $^{36}$Ca, $^{36}$S, $^{38}$Ca, and $^{38}$Ar.

\begin{figure}[htb]
\includegraphics[width=0.82\linewidth, clip, trim = 10 20 10 20]{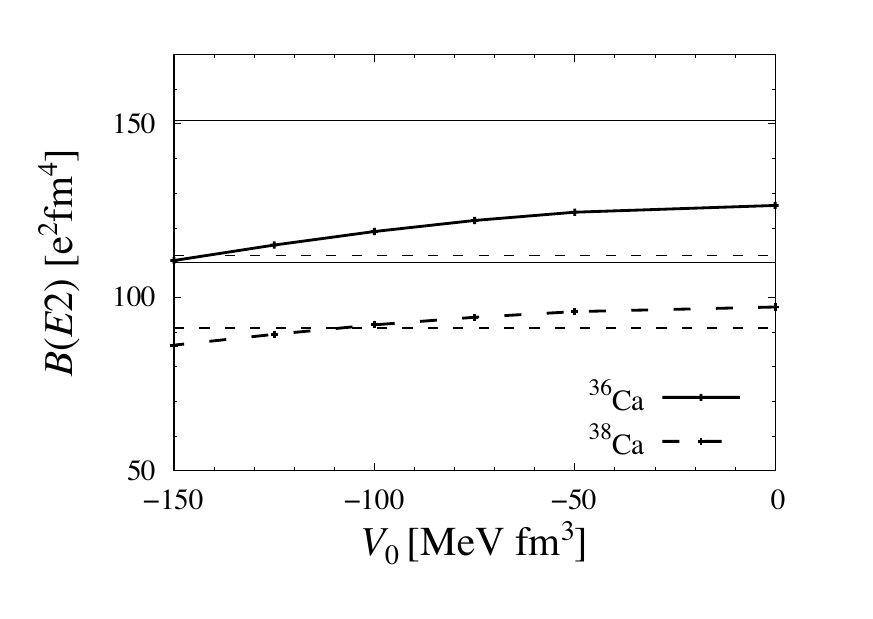}

\includegraphics[width=0.82\linewidth, clip, trim = 10 20 10 20]{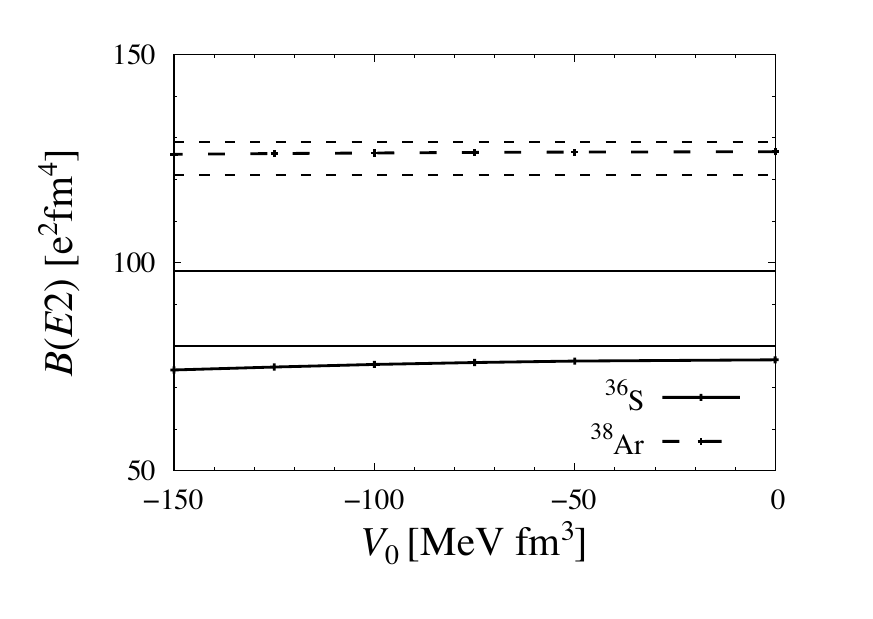}

\caption{Upper panel: The  $B(E2\uparrow)$ transition probabilities  in $^{36}$Ca (the solid line) and $^{38}$Ca (the long-dashed line), calculated in SMEC, are plotted as a function of the continuum coupling strength $V_0$. Horizontal solid and short-dashed lines show the uncertainties of the experimental $B(E2\uparrow)$ values in $^{36}$Ca and $^{38}$Ca, respectively. The $T=1$ cross-shell monopoles of ZBM-IO interaction in this panel are: ${\cal M}^{T=1}(1s_{1/2}\,1p_{3/2})=-2.477$~MeV and ${\cal M}^{T=1}(1s_{1/2}\,0f_{7/2})=-2.177$~MeV.  \\
Lower panel: The same as in the upper panel but for the mirror nuclei $^{36}$S and $^{38}$Ar. The $T=1$ cross-shell monopoles in this case are: ${\cal M}^{T=1}(1s_{1/2}\,1p_{3/2})=-0.077$~MeV and ${\cal M}^{T=1}(1s_{1/2}\,0f_{7/2})=-2.177$~MeV. 
}
\label{be2mono}
\end{figure}
Dependence of the $B(E2,0^+_1 \rightarrow 2^+_1)$ transitions probabilities on the continuum coupling strength $V_0$ is shown in Fig.~\ref{be2mono} for mirror pairs of nuclei ($^{36}$Ca, $^{36}$S), and ($^{38}$Ca, $^{38}$Ar). The cross-shell monopoles in ZBM-IO interaction are kept constant as a function of $V_0$. Horizontal lines show the experimental uncertainties of $B(E2\uparrow)$. For more information, see the caption of Fig.~\ref{be2mono}. One may notice a weak dependence on $V_0$ in $^{36}$Ca. Even weaker dependence is seen in $^{38}$Ca, and almost no dependence on the strength of the continuum coupling is in $^{38}$Ar. The $B(E2\uparrow)$ curves shown in this figure have been obtained for the effective charges equal: $e_p=1.236$, $e_n=0.409$.
 
\begin{figure}[htb]
\includegraphics[width=0.82\linewidth, clip, trim = 10 20 10 20]{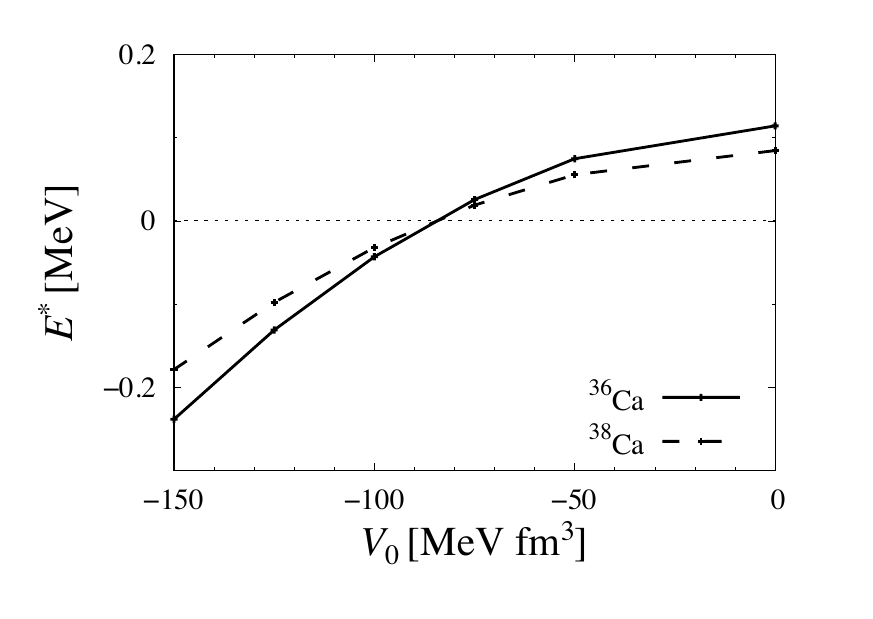}

\includegraphics[width=0.82\linewidth, clip, trim = 10 20 10 20]{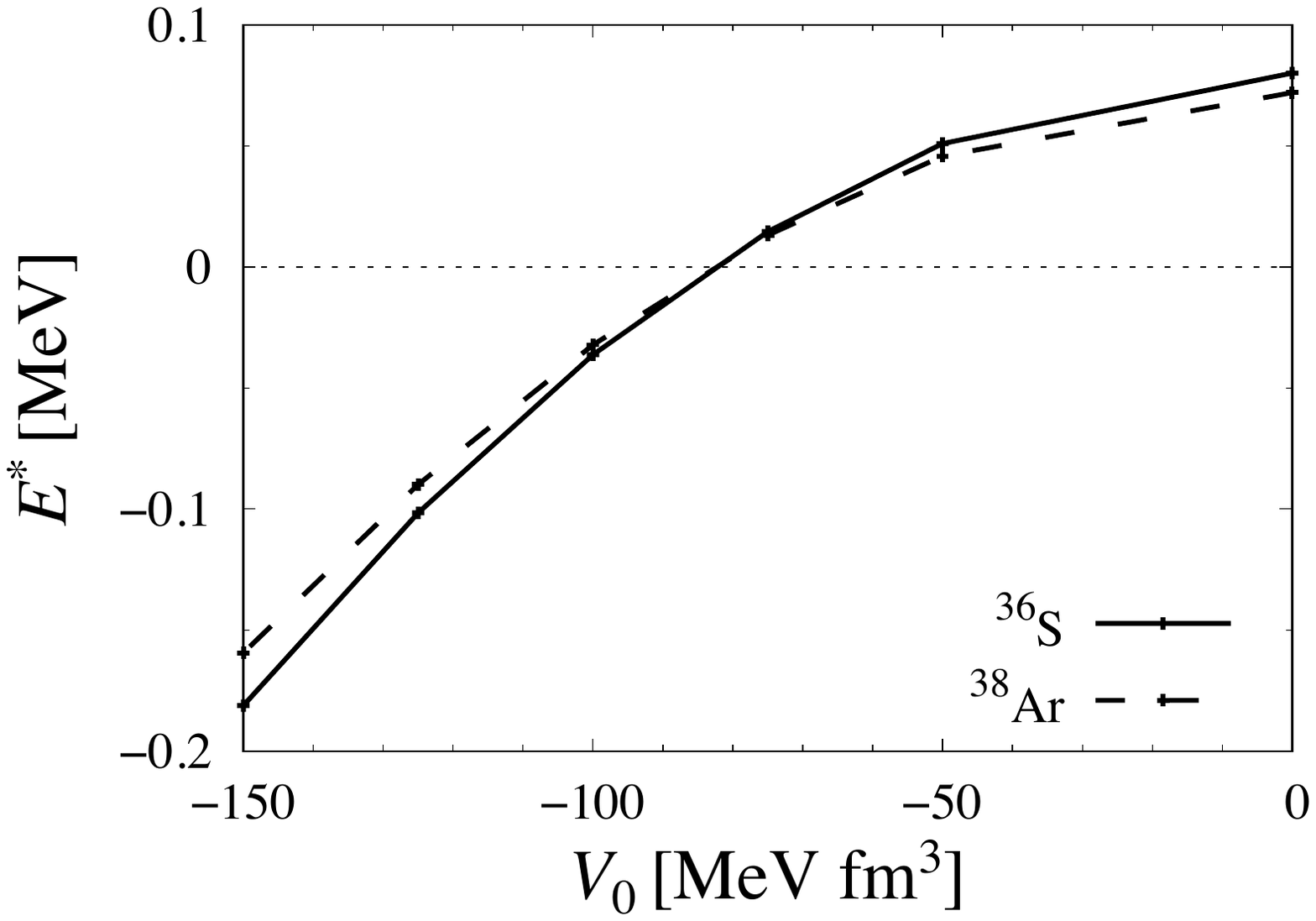}

\caption{Upper panel: Ground state energy of $^{36}$Ca and $^{38}$Ca is plotted as a function of the continuum coupling strength $V_0$ for fixed values of $T=1$ cross-shell monopoles: ${\cal M}^{T=1}(1s_{1/2}\,1p_{3/2})=-2.477$~MeV and ${\cal M}^{T=1}(1s_{1/2}\,0f_{7/2})=-2.177$~MeV. The dashed horizontal line corresponds to the situation when the ground state $0^+_1$ is at the experimental distance from the state $2^+_1$.\\
Lower panel: The same as in the upper panel but for mirror nuclei $^{36}$S and $^{38}$Ar. The $T=1$ cross-shell monopoles in this case are: ${\cal M}^{T=1}(1s_{1/2}\,1p_{3/2})=-0.077$~MeV and ${\cal M}^{T=1}(1s_{1/2}\,0f_{7/2})=-2.177$~MeV. 
}
\label{zeroener}
\end{figure}
Dependence of the ground state energy of $^{36}$Ca, $^{38}$Ca, $^{36}$S and $^{38}$Ar on the continuum-coupling strength parameter $V_0$ is shown in Fig.~\ref{zeroener}. One can see that this dependence is rather weak for all considered nuclei. The strongest effect  is seen in the ground state of $^{36}$Ca. The change of an excitation energy of the $2^+_1$ state is mainly due to the down-sloping of the ground state energy in all considered nuclei.

\begin{figure}
\includegraphics[width=0.9\linewidth, clip, trim = 10 20 10 0]{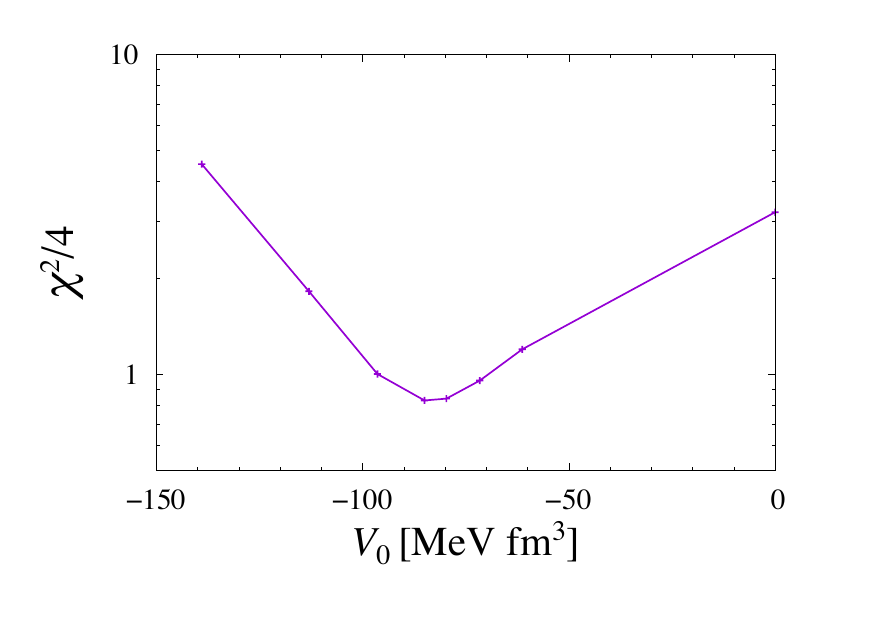}
\caption{Function $\chi^2$ for calculated $B(E2\uparrow)$ values in $^{36}$Ca, $^{38}$Ca, $^{36}$S and $^{38}$Ar (see Tables~\ref{tab:smecres} and \ref{tab:smecres-mir}) is plotted as a function of the continuum coupling strength $V_0$ and normalized to the number of data. At the minimum, value of the $\chi^2$ function is: $\chi^2/4=0.82$. 
}
\label{smecsol-chi-al}
\end{figure}
Using the calculated $B(E2\uparrow)$ values for different continuum-coupling strengths $V_0$ (see the Tables~\ref{tab:smecres} and \ref{tab:smecres-mir}), one can search for optimal parameters of the ZBM-IO interaction  which reproduce best the experimental $B(E2\uparrow)$ values. Figure~\ref{smecsol-chi-al} presents the function $\chi^2$ normalized to the number of data points which is plotted as a function of $V_0$ for all $B(E2\uparrow)$ values. We can see that the preferable continuum coupling strength is $V_0\approx -85$~MeV~fm$^3$. 

\begin{table}[htb]
    \caption{Results of the $\chi^2$ analysis of $B(E2\uparrow)(V_0)$ (see Fig.~\ref{smecsol-chi-al} and Tables \ref{tab:smecres}, \ref{tab:smecres-mir}) for $^{36}$Ca, $^{38}$Ca, $^{36}$S and $^{38}$Ar. Theoretical errors are due to experimental uncertainties of the separation energies~\cite{ensdf}. $B(E2\uparrow)$ are given in units $e^2$fm$^4$, monopole terms are in MeV, and $V_0$ is given in units MeV fm$^3$. The effective charges are: $e_p=1.236$, $e_n=0.409$.  Experimental $B(E2\uparrow)$ values for $^{36,38}$Ca are taken from Ref.~\cite{PhysRevC.107.034306} and for $^{36}$S, $^{38}$Ar from Ref.~\cite{PRITYCHENKO201773}}.
    \vskip 0.3truecm
    \begin{tabular}{c|cc|cc}
    \hline
    \hline
  & $^{36}$Ca  &  $^{38}$Ca  &  $^{36}$S  &  $^{38}$Ar \\
    \hline
 $1s_{1/2}\,0f_{7/2}$ & \multicolumn{2}{c|}{-2.177} & \multicolumn{2}{c}{-2.167}  \\
 $1s_{1/2}\,1p_{3/2}$ & \multicolumn{2}{c|}{-2.477} & \multicolumn{2}{c}{-0.077}  \\
 $V_0$ & $-85^{+17}_{-14}$ & $-85.16^{+0.11}_{-0.12}$  & $-82.98\pm 0.10$ & $-83.01\pm 0.12$ \\
 $B(E2\uparrow)$  & $120.9^{+1.9}_{-1.8}$ & $93.375\pm 0.01$ & $75.826\pm 0.002$ & $126.4190\pm 0.0005$ \\
 \hline
  exp.: & $131\pm 20$ & $101\pm 11$ & $89\pm 9$ & $125\pm 4$ \\
  \hline
  \hline
   \end{tabular}
  \label{tab:final}
\end{table}
SMEC results corresponding to the minimum of $\chi^2$ function (see Fig.~\ref{smecsol-chi-al}) are presented in Table~\ref{tab:final}. The optimal monopole 
${\cal M}^{T=1}(1s_{1/2}, 0f_{7/2})\simeq -2.17$ MeV is almost the same in mirror pairs ($^{36}$Ca/$^{36}$S) and ($^{38}$Ca/$^{38}$Ar). On the other hand, the optimal monopole ${\cal M}^{T=1}(1s_{1/2}, 1p_{3/2})$ changes strongly going from $^{36,38}$Ca to their mirror partners $^{36}$S,$^{38}$Ar. The optimal continuum-coupling strength is almost the same in $^{36,38}$Ca, $^{36}$S, $^{38}$Ar and equals $V_0\simeq-83$ MeV fm$^3$.

\begin{table}[htb]
    \caption{$F_n(0)$, $F_n(2)$, $F_p(0)$ and $F_p(2)$ for $0^+_1$ and $2^+_1$ states in all considered nuclei for the most favorable monopole modifications ({\it cf} Table~\ref{tab:final}).}
    \vskip 0.3truecm
    \begin{center}
    \begin{tabular}{cccccc}
    \hline
    \hline
Nucleus & $J^\pi$ & $F_p(0)$  &  $F_p(2)$  &  $F_n(0)$  &  $F_n(2)$ \\
\hline
 \multirow{2}{*}{$^{36}$Ca} 
& $0^+$  &  0.6157 & 0.3664 & 0.9792 & 0.0029 \\
& $2^+$  &  0.1295 & 0.8588 & 0.9880 & 0.0003 \\
 \multirow{2}{*}{$^{38}$Ca}
& $0^+$  &  0.5542 & 0.3639 & 0.8744 & 0.0436 \\
& $2^+$  &  0.3413 & 0.5770 & 0.9117 & 0.0066 \\

\hline
  \multirow{2}{*}{$^{36}$S}
& $0^+$  &  0.9842 & 0.0030 & 0.6695 & 0.3177 \\
& $2^+$  &  0.9829 & 0.0005 & 0.2575 & 0.7259 \\
  \multirow{2}{*}{$^{38}$Ar}
& $0^+$  &  0.8858 & 0.0415 & 0.5591 & 0.3683 \\
& $2^+$  &  0.9223 & 0.0055 & 0.3202 & 0.6076 \\
    \hline
    \hline
   \end{tabular}
    \end{center}
  \label{tab:Fqmirror}
\end{table}
In Table~\ref{tab:Fqmirror} we have collected $F_q$ values for the most favorable monopole modification (see Table~\ref{tab:final}) in $^{36,38}$Ca,$^{36}$S, and $^{38}$Ar. For the ground state $0^+_1$ in $^{36}$Ca, fraction of the $F_p(0)$ part excited from $(1s_{1/2} 0d_{3/2})$ to $(0f_{7/2} 1p_{3/2})$ is close to the fraction $F_n(0)$ in the mirror nucleus $^{36}$S, also the fractions $F_n(0)$ in $^{36}$Ca and $F_p(0)$ in $^{36}$S are nearly identical. In $^{38}$Ca, fraction of the proton part $F_p(0)$ is almost identical with the fraction of the neutron part $F_n(0)$ in $^{38}$Ar. Similarly the fraction $F_n(0)$ in $^{38}$Ca is practically identical with the $F_p(0)$ fraction in $^{38}$Ar. Therefore, one may conclude that whereas the mirror symmetry in the ground state of nuclei $^{36}$Ca and $^{36}$S is only slightly broken, so it is satisfied in $^{38}$Ca and $^{38}$Ar.  

A different situation is seen in the $2^+_1$ state. The fraction of neutrons excited from $(1s_{1/2} 0d_{3/2})$ to $(0f_{7/2} 1p_{3/2})$ in $^{36}$S is significantly smaller than the corresponding fraction of the proton part in $^{36}$Ca. This striking difference in the occupation of $0f_{7/2} 1p_{3/2}$ shells has also its counterpart in the reverse order of $2^+_1$ and $0^+_2$  states in $^{36}$Ca as compared with $^{36}$S~\cite{PhysRevLett.129.122501}. In $^{38}$Ca, $^{38}$Ar, on the contrary, the mirror symmetry in $2^+_1$ states is well preserved.

Additional information about the structure of mirror pairs of nuclei: $^{36}$Ca, $^{36}$S and $^{38}$Ca, $^{38}$Ar, is provided by the spectroscopic factors presented in Table~\ref{tab:specfac}. One may notice that spectroscopic factors in the ground states of all considered nuclei are significantly larger than in the excited states $2^+_1$ and $0^+_2$. In particular,
$C^2{\cal S}_{s_{1/2}}(0^+_2)$ and  $C^2{\cal S}_{s_{1/2}}(2^+_1)$ 
are smaller  by one and three orders of magnitude, respectively, than the ground state spectroscopic factor $C^2{\cal S}_{s_{1/2}}(0^+_1)$. Moreover, the spectroscopic factor $C^2{\cal S}_{s_{1/2}}(0^+_1)$ in $^{36}$Ca, $^{36}$S is larger than the corresponding spectroscopic factor in $^{38}$Ca, $^{38}$Ar. 

\begin{table}[htb]
    \caption{SMEC spectroscopic factors  for $0^+_1$, $0^+_2$, $2^+_1$ states in $^{36}$Ca, $^{38}$Ca, $^{36}$S and $^{38}$Ar. ${\cal C}^2{\cal S}_{\ell_j}(J^{\pi}_i)$ denote proton (neutron) spectroscopic factors in $^{36,38}$Ca ($^{36}$S,$^{38}$Ar).}
    \vskip 0.3truecm
    \begin{center}
    \begin{tabular}{ccccc}
    \hline
    \hline  \\[-10pt]
Nucleus & $C^2{\cal S}_{s_{1/2}}(0^+_1)$  &  $C^2{\cal S}_{s_{1/2}}(0^+_2)$ & $C^2{\cal S}_{d_{3/2}}(2^+_1)$  &  $C^2{\cal S}_{s_{1/2}}(2^+_1)$ \\[2pt]
\hline
 $^{36}$Ca 
& 2.903 & 0.534 & 0.0030 & 0.0054 \\
 $^{38}$Ca
& 2.320 & 0.586 & 0.0408 & 0.0078 \\ 
  $^{36}$S
& 3.019 & 0.470 & 0.0004 & 0.0070 \\
  $^{38}$Ar
& 2.300 & 0.611 & 0.0368 & 0.0080 \\    
    \hline
    \hline
   \end{tabular}
    \end{center}
  \label{tab:specfac}
\end{table}
Finally two observations are common. First, the excitation energy of $2^+_1$ always increases with increasing the continuum coupling strength even though this state couples only weakly to the continuum due to a very small spectroscopic factor $C^2{\cal S}_{s_{1/2}}(2^+_1)$. This is due to a much stronger influence of the coupling to the continuum on the ground states, even though these states are well bound. Secondly, $B(E2\uparrow)$ always decreases with increasing continuum coupling strength for any combinations of monopole changes. 

The mirror symmetry in the pair of nuclei $^{38}$Ca and $^{38}$Ar is well satisfied as the spectroscopic factors are almost identical (see Table~\ref{tab:specfac}), and the difference of ground state energies as well as the difference of $2^+_1$ energies in these nuclei is very small. 

On the contrary, the mirror symmetry in $^{36}$Ca and $^{36}$S is strongly violated. The spectroscopic factors $C^2{\cal S}_{s_{1/2}}(0^+_1)$, $C^2{\cal S}_{s_{1/2}}(0^+_2)$ in this pair of nuclei are different. Moreover, the order of $2^+_1$ and $0^+_2$ states is different in $^{36}$Ca and in $^{36}$S. This effect can be traced back partially to the proximity of the proton decay threshold in $^{36}$Ca and to the larger spectroscopic factor $C^2{\cal S}_{s_{1/2}}(0^+_2)$ in $^{36}$Ca than in $^{36}$S.

In conclusion, anomalous values of $B(E2,0^+_1 \rightarrow 2^+_1)$ transition probability in $^{36}$Ca and $^{38}$Ca as compared with the values in the mirror nuclei $^{36}$S and $^{38}$Ar, revealed strong mirror symmetry breaking in the pair $^{36}$Ca and $^{36}$S. The SMEC provides good description of all four $B(E2,0^+_1 \rightarrow 2^+_1)$ transition probabilities using the ZBM-IO interaction with a monopole term ${\cal M}^{T=1}(1s_{1/2},1p_{3/2})$ which is modified when going from $^{36,38}$Ca to $^{36}$S, $^{38}$Ar.

The most important difference is seen in the structure of $2^+_1$ state in $^{36}$Ca and $^{36}$S. The fraction $F_p(2)$ of proton part excited from $(1s_{1/2}0d_{3/2})$ to $(0f_{7/2}1p_{3/2})$ in the proton unbound $2^+_1$ state of $^{36}$Ca is considerably larger than the corresponding neutron part $F_n(2)$ in $^{36}$S. In general, main effects of the continuum coupling in studied nuclei are concentrated in the ground state $0^+_1$ and to a smaller extent in the excited $0^+_2$ state. The reverse order of the $2^+_1$ and $0^+_2$ states in $^{36}$Ca is the combined effect of the proximity of $0^+_2$ resonance to the particle emission threshold and the spectroscopic factor ${\cal C}^2{\cal S}_{s_{1/2}}(0^+_2)$ which is two-orders of magnitude bigger than the ${\cal C}^2{\cal S}_{s_{1/2}}(2^+_1)$.

\vskip 1truecm
\textit{Acknowledgements--}
We wish to thank Witek Nazarewicz and Simin Wang for a useful discussion.

\bibliography{refs}

\begin{thebibliography}{10}

\bibitem{PhysRevC.107.034306}
N.~Dronchi, D.~Weisshaar, B.~A. Brown, A.~Gade, R.~J. Charity, L.~G. Sobotka,
  K.~W. Brown, W.~Reviol, D.~Bazin, P.~J. Farris, A.~M. Hill, J.~Li,
  B.~Longfellow, D.~Rhodes, S.~N. Paneru, S.~A. Gillespie, A.~Anthony,
  E.~Rubino, and S.~Biswas.
\newblock Measurement of the $b(e2\ensuremath{\uparrow})$ strengths of
  $^{36}\mathrm{Ca}$ and $^{38}\mathrm{Ca}$.
\newblock {\em Phys. Rev. C}, 107:034306, Mar 2023.

\bibitem{PhysRevC.108.L061301}
T.~Beck, A.~Gade, B.~A. Brown, J.~A. Tostevin, D.~Weisshaar, D.~Bazin, K.~W.
  Brown, R.~J. Charity, P.~J. Farris, S.~A. Gillespie, A.~M. Hill, J.~Li,
  B.~Longfellow, W.~Reviol, and D.~Rhodes.
\newblock Probing proton cross-shell excitations through the two-neutron
  removal from $^{38}\mathrm{Ca}$.
\newblock {\em Phys. Rev. C}, 108:L061301, Dec 2023.

\bibitem{PhysRevLett.100.152502}
C.~R. Hoffman, T.~Baumann, D.~Bazin, J.~Brown, G.~Christian, P.~A. DeYoung,
  J.~E. Finck, N.~Frank, J.~Hinnefeld, R.~Howes, P.~Mears, E.~Mosby, S.~Mosby,
  J.~Reith, B.~Rizzo, W.~F. Rogers, G.~Peaslee, W.~A. Peters, A.~Schiller,
  M.~J. Scott, S.~L. Tabor, M.~Thoennessen, P.~J. Voss, and T.~Williams.
\newblock Determination of the $n=16$ shell closure at the oxygen drip line.
\newblock {\em Phys. Rev. Lett.}, 100:152502, Apr 2008.

\bibitem{APP2008}
J.~Okolowicz, M.~Ploszajczak, and Yan an~Luo.
\newblock Continuum coupling effects in spectra of mirror nuclei and binding
  systematics.
\newblock {\em Acta Physica Polonica B}, 39:389--400, 2007.

\bibitem{399d5bb09b5d11dcbee902004c4f4f50}
A.~Buerger, F.~Azaiez, Z.~Dombradi, A.~Alogra, B.~Bastin, G.~Benzoni,
  A.~Al-Kahtib, E.~Clement, Z~Dlouhy, A.~Gorgen, S.~Grevy, W.~Korten, Geirr
  Sletten, C.~Timis, D.~Verney, and S.~Williams.
\newblock Spectroscopy around 36ca.
\newblock {\em Acta Physica Polonica B}, 38(4):1353--1357, 2007.

\bibitem{PhysRevLett.129.122501}
L.~Lalanne, O.~Sorlin, A.~Poves, M.~Assi\'e, F.~Hammache, S.~Koyama, D.~Suzuki,
  F.~Flavigny, V.~Girard-Alcindor, A.~Lemasson, A.~Matta, T.~Roger, D.~Beaumel,
  Y.~Blumenfeld, B.~A. Brown, F.~De~Oliveira Santos, F.~Delaunay,
  N.~de~S\'er\'eville, S.~Franchoo, J.~Gibelin, J.~Guillot, O.~Kamalou,
  N.~Kitamura, V.~Lapoux, B.~Mauss, P.~Morfouace, M.~Niikura, J.~Pancin, T.~Y.
  Saito, C.~Stodel, and J-C. Thomas.
\newblock Structure of $^{36}\mathrm{Ca}$ under the coulomb magnifying glass.
\newblock {\em Phys. Rev. Lett.}, 129:122501, Sep 2022.

\bibitem{PRITYCHENKO201773}
B.~Pritychenko, M.~Birch, and B.~Singh.
\newblock Revisiting grodzins systematics of b(e2) values.
\newblock {\em Nuclear Physics A}, 962:73--102, 2017.

\bibitem{Okolowicz2003}
J.~Oko{\l}owicz, M.~P{\l}oszajczak, and I.~Rotter.
\newblock Dynamics of quantum systems embedded in a continuum.
\newblock {\em Phys. Rep.}, 374(4):271 -- 383, 2003.

\bibitem{bennaceur_2000}
K.~Bennaceur, F.~Nowacki, J.~Oko{\l}owicz, and M.~P{\l}oszajczak.
\newblock {\em Nucl. Phys. A}, 671:203, 2000.

\bibitem{rotureau_2006}
J.~Rotureau, J.~Oko{\l}owicz, and M.~P{\l}oszajczak.
\newblock {\em Nucl. Phys. A}, 767:13, 2006.

\bibitem{CAURIER2001240}
E~Caurier, K~Langanke, G~Martínez-Pinedo, F~Nowacki, and P~Vogel.
\newblock Shell model description of isotope shifts in calcium.
\newblock {\em Physics Letters B}, 522(3):240--244, 2001.

\bibitem{PhysRevC.63.044316}
S.~Nummela, P.~Baumann, E.~Caurier, P.~Dessagne, A.~Jokinen, A.~Knipper,
  G.~Le~Scornet, C.~Mieh\'e, F.~Nowacki, M.~Oinonen, Z.~Radivojevic,
  M.~Ramdhane, G.~Walter, and J.~\"Ayst\"o.
\newblock Spectroscopy of ${}^{34,35}\mathrm{Si}$ by $\ensuremath{\beta}$
  decay: $sd\ensuremath{-}fp$ shell gap and single-particle states.
\newblock {\em Phys. Rev. C}, 63:044316, Mar 2001.

\bibitem{ensdf}
\url{http://www.nndc.bnl.gov/ensdf}, 2015.

\end{thebibliography}

\end{document}